\documentclass{article}
\usepackage{spconf,amsmath,amssymb,graphicx}
\usepackage{multirow}
\usepackage{hhline}
\usepackage{romannum}
\usepackage{fixltx2e}
\usepackage{xcolor}
\usepackage{pgfplots}
\usepackage{tikz}
\DeclareMathOperator{\E}{\mathbb{E}}

\title{UTD-CRSS submission for MGB-3 Arabic Dialect Identification: \\ Front-end and Back-end Advancements on Broadcast Speech}
\name{Ahmet E. Bulut, Qian Zhang, Chunlei Zhang, Fahimeh Bahmaninezhad, John H. L. Hansen\thanks{This project was funded by AFRL under contract FA8750-15-1-0205 and partially by the University of Texas at Dallas from the Distinguished University Chair in Telecommunications Engineering held by J. H. L. Hansen.}}
\address{Center for Robust Speech Systems (CRSS) \\
University of Texas at Dallas, Richardson, TX 75080 \\
{\small \tt \{ahmet.bulut, qian.zhang, chunlei.zhang, fahimeh.bahmaninezhad, john.hansen\}@utdallas.edu}}

\begin{document}
%
\maketitle
\begin{abstract}

This study presents systems submitted by the University of Texas at Dallas, Center for Robust Speech Systems (UTD-CRSS) to the MGB-3 Arabic Dialect Identification (ADI) subtask. This task is defined to discriminate between five dialects of Arabic, including Egyptian, Gulf, Levantine, North African, and Modern Standard Arabic. We develop multiple single systems with different front-end representations and back-end classifiers. At the front-end level, feature extraction methods such as Mel-frequency cepstral coefficients (MFCCs) and two types of bottleneck features (BNF) are studied for an i-Vector framework. As for the back-end level, Gaussian back-end (GB), and Generative Adversarial Networks (GANs) classifiers are applied alternately. The best submission (contrastive) is achieved for the ADI subtask with an accuracy of 76.94\% by augmenting the randomly chosen part of the development dataset. Further, with a post evaluation correction in the submitted system, final accuracy is increased to 79.76\%, which represents the best performance achieved so far for the challenge on the test dataset. 


\end{abstract}
\section{Introduction}
\label{sec:intro}

Besides Language Identification (LID), the task of Dialect Identification (DID) has recently become more of an issue especially given the wider diversity of spoken languages. DID can be defined as a sub-task of LID while determining the pattern of pronunciation and/or grammar of a language used within some geographical region by the community of native speakers \cite{lei2011adiGMM}. Identifying such intra-language varieties have a great importance for a number of language-related speech processing tasks. One such task is Automatic Speech Recognition (ASR), in which the effect of dialect/accent has been investigated and noted that it is one of the most important factors that influence ASR performance \cite{gupta1982effects, Huang20011377}. Research has shown that \cite{diakoloukas1997development, humphries1998use, liu2000mandarin, ward2002lexicon, soltau2011modern} ASR systems may benefit from effective DID by applying dialect specific training and adaptation.

Arabic is one of the most widely spoken languages worldwide. Within the area where Arabic is spoken, there are four generally accepted dialects: Egyptian (EGY), Gulf (GLF), Levantine (LEV), North African (NOR). In addition to these four dialects, Modern Standard Arabic (MSA) is considered as the standard written and spoken language throughout the Arab world. In general, MSA is not a native language of any specific Arabic speaking people, but is generally used in news media, speeches, and academic publishing \cite{chiang2006parsing}. However, MSA is considered as one of the general dialects in Arabic based speech technology. As the study \cite{AliDCKYG0R16} shows, very basic colloquial sentences can include lexical variations across the different dialects. The Arabic Dialect Identification (ADI) task can be regarded as LID task when compared to the other languages. Therefore, a wide range of acoustic and lexical features are exploited in the literature for ADI systems.

In one recent study of ADI \cite{Zaidan2014ADI}, it composed a large amount of text-based Arabic data and set up an online dialect annotation system. The authors trained various word and letter based models in order to classify dialectal sentences. In another domain, for improved machine translation an ADI system was developed with textual social media data by using both supervised and semi-supervised classification methods \cite{huang2015improved}. As for acoustic based systems, Biadsy et al. developed an ADI system using phonotactic modeling \cite{Biadsy2009SAD}. Gaussian mixture models were used for another study \cite{lei2011adiGMM} which achieved competitive results with a modest amount of training data. In order to leverage the contribution of complementary features, \cite{hansen2016unsupervised} combined lexical based systems with i-Vector based acoustic systems and achieved superior results. They used logistic regression and GB classifiers for their text and acoustic based systems, respectively. This study \cite{Boril2012} analyzed the non-linguistic content of the DID problem and drawn attention to the channel characteristics of target database. In a recent challenge which includes an ADI shared task was organized as part of the Vardial Evaluation Campaign 2017 \cite{zampieri2017vardial17}. Many systems in the challenge incorporated word and string based systems with a system using bottleneck features (BNF) and i-Vector models \cite{AliDCKYG0R16}. Moreover, neural networks based methods were employed by some participants \cite{belinkov2016character,guggilla2016discrimination} for the task, but they were considered of limited use due to the amount of training data provided \cite{malmasi2016discriminating}. In our proposed system, we combine five individual systems. Four are acoustic based systems and one is a lexical based system. In the lexical based system, we basically use unigram term frequency features followed by an SVM classifier \cite{yu2013libshorttext} provided through ASR output transcripts \cite{AliDCKYG0R16}. For the first two acoustic-based systems, we use i-Vector features extracted by using MFCC features and apply two alternate classifiers namely GB \cite{gang_ICASSP13_backend} and GANs \cite{salimans2016improved}, respectively. In our third acoustic based system, we use BNF-based i-Vectors \cite{AliDCKYG0R16} following by GB classifier. Finally, in our last system, we use unsupervised bottleneck features (UBNF) \cite{zhang2017ubnf} for i-Vector extraction and classify with GANs. The described combination of the system submission has the best performance on the MGB-3 ADI test data between the sites participated both Vardial 2017 \cite{zampieri2017vardial17} and MGB-3 ADI \cite{Ali2017mgb3} tasks.

The remainder of this paper is organized as follows. In the next section, we present the theoretical details of the proposed systems. Sec. \ref{sec:database} includes the description  of  dialect  corpora  and their  statistics,  followed  by  a  baseline  system  description  in Sec. \ref{sec:basesys}. The details of the conducted experiments and results are presented in Sec. \ref{sec:exps&results} and Sec. \ref{sec:submissions} describe the details of the submitted systems. The paper is concluded in Sec. \ref{sec:conclude} with a brief summary of the results. 

\section{Proposed Dialect Classification Systems}
\label{sec:propsys}

In order to benefit from the complementary power of text and acoustic based systems, we develop a number of systems for both domains. The following sub-sections present text and acoustic based systems which are investigated in detail.

\subsection{Text-based Systems}
\label{ssec:textsys}

With the assumption that each dialect should demonstrate some unique word patterns, several popular NLP approaches are employed for many DID tasks. As stated in Sec.\ref{sec:intro}, providing ASR transcripts are used for text based system. Next, we briefly describe the pre-processing methods, feature generation strategies, and back-end classifiers we apply in each part by using an open source software \cite{yu2013libshorttext}.

\subsubsection{Text pre-processing}
\label{sssec:textpreproc}

For a given language and dialect, there are many words that bear little or no meaning, but are necessary for correct grammatical structure. Those words are defined as ``stop-words". For example, in the English language words such as: ``a", ``the", ``and" have very little meaning but are frequently used in the language \cite{greenberg1997origins}. 

Another text pre-processing method is defined as stemming, which means reducing the derived words to their word stem. For example, if an English document would contain two instances of the word ``study", two instances of the word ``studies", and one instance of the word ``studying", then a stemming reduction for those three separate word features would reduce them to one word that determines the number of the set of words with the root ``study" as occurred in the document five times \cite{hansen2016unsupervised}.

After applying the necessary text pre-processing methods, we perform an n-grams procedure. In order to understand the n-grams model, let assume each sentence can be regarded as a sequence of words $W=\{w_1, w_2, \ldots , w_n\}$ which includes a set of $n$ words. The probability of generating the word sequence $W$ can be measured as,

\vspace*{-5pt}
\begin{eqnarray}
\label{eq:ngram}
\begin{split}
P(W) &= P(w_1, w_2, \ldots, w_n) \\
     &= P(w_1)P(w_2|w_1) \ldots P(w_n|w_1, \ldots, w_{n-1}) \\
     &= \prod_{i=1}^{n} P(w_i|w_1, \ldots, w_{i-1}).
\end{split}
\end{eqnarray}

\noindent with unigram and bigram models, we basically assume that the probability of a given word sequence defined in Eq. \ref{eq:ngram}, would reduce to Eq. \ref{eq:unigram} and Eq. \ref{eq:bigram}, respectively.

\vspace*{-5pt}
\begin{eqnarray}
P(W) &\approx& \prod_{i=1}^{n} P(w_i) \label{eq:unigram}, \\
P(W) &\approx& \prod_{i=1}^{n} P(w_i|w_{i-1}). \label{eq:bigram}
\end{eqnarray}

\noindent Because data scarcity for the ADI task, we only investigate unigram and bigram models for the proposed text based systems. 

The collection of all pre-processing procedures are used sequentially with combinations listed in Table \ref{tab:txpreproc}.

\vspace{-6pt}
\begin{table}[h]\caption{Combination of ADI text pre-processing methods. Second and third columns indicate application of stop-word removal and stemming, where the fourth column indicates `Y': bigram and `N': unigram.}\label{tab:txpreproc}
\vspace{5pt}
\centering
\scalebox{0.90}{
\begin{tabular}{|c c c c|}
\hline
\textbf{Comb. \#} & \textbf{Stop-word} & \textbf{Stemming} & \textbf{Bigram} \\ \hline \hline
0 & N & N & N \\ 
1 & N & N & Y \\ 
2 & N & Y & N \\ 
3 & N & Y & Y \\ 
4 & Y & N & N \\ 
5 & Y & N & Y \\ 
6 & Y & Y & N \\ 
7 & Y & Y & Y \\ \hline
\end{tabular}
}
\end{table}

\subsubsection{Text Feature Generation \& Classification}
\label{sssec:textfeatgen}

Once the raw text ASR transcripts are pre-processed, among the several alternative feature extraction approaches, Binary (BIN), Term Frequency (TF), and Term Frequency-Inverse Document Frequency (TF-IDF) features are applied within the scope of this study. Basically, BIN features assign either 0 or 1 to represent whether some word/word-pairs are present or not, and TF features signify how often a word/word-pair occurs in a document. Finally, TF-IDF features represent how important a word/word-pair is to a document in a collection.

For the back-end classification of the text based systems, we use a multi-class SVM similar to the baseline text based system. 

\subsection{Acoustic-based Systems}
\label{ssec:acusys}
In this section, we present details of the proposed acoustic based systems. By exploiting three different features and two different classifiers, four acoustic based system developed for the ADI task.

\subsubsection{Features}
\label{sssec:feats}

MFCC features are considered as raw features and have been used extensively in the speech processing community for several problems. Furthermore, many systems convert MFCC features into more discriminating features such as i-Vector features \cite{dehak2011front}. Therefore, MFCC i-Vectors are the first feature type used in our systems.

\begin{figure}[htb]
\caption{The UBNF feature extraction diagram \cite{zhang2017ubnf}.}
\vspace{5pt}
\centerline{\includegraphics[width=8.5cm]{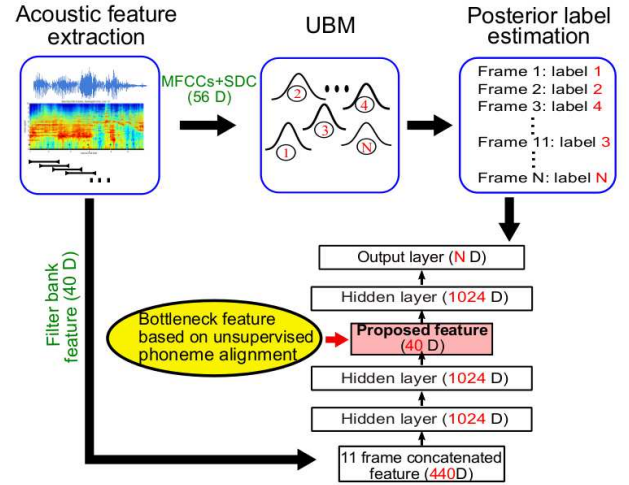}}
\label{fig:bottleneck}
\end{figure}

Traditional bottleneck features have recently become popular as an alternative to MFCCs for several speech tasks since they contain both acoustic and phonetic information. We use the BNF i-Vector features which are provided by the challenge organizers. However, since the transcripts for the corresponding audio files are ASR outputs, we apply UBNF i-Vectors in order to capture much more discriminating information. In a previous study \cite{zhang2017ubnf}, unsupervised bottleneck features (UBNF) were successfully applied for the DID task and shown to be more effective for an i-Vector framework compared to MFCCs and traditional BNFs. The basic concept is similar to traditional bottleneck feature but without requiring an extra transcribed training corpus. Instead of forced aligned senone labels, the GMM mixture numbers assumed to represent phonetic information. The details of UBNF feature extraction diagram is shown in Fig. \ref{fig:bottleneck}.

\subsubsection{Back-end Classifiers}
\label{sssec:classify}

\textbf{(a) Gaussian Back-end:} The generative GB which has generally been used for the LID and DID tasks is the first classifier that we employ. For each dialect, the distribution of i-Vectors was modeled by a Gaussian distribution by assuming a common full covariance matrix for all dialects. For each i-Vector, $w$ stands for a test utterance, and we calculate the log-likelihood for each dialect as:


\vspace*{-5pt}
\begin{eqnarray}
\label{eq:loglike}
\begin{split}
\log p(w|d) = &-\dfrac{1}{2}w^T \Sigma^{-1} w + w^T \Sigma^{-1} m_d \\
&- \dfrac{1}{2} m_d ^T \Sigma^{-1} m_d +c,
\end{split}
\end{eqnarray}

\noindent where $m_d$ is the mean vector for the dialect $d$, $\Sigma$ is the shared covariance matrix, and $c$ is a constant. 
 
\noindent \textbf{(b) GANs Back-end:} As a second back-end classifier, we use Generative Adversarial Networks (GANs) which is originally proposed in \cite{goodfellow2014generative}. To our best knowledge, this is the first time that GANs is used as a back-end classifier for a DID task.

We incorporate the concept of semi-supervised adversarial training for dialect identification, given the unlabeled data for the MGB-3 ADI challenge \cite{zhang2017letters}. Instead of using discriminator network to detect whether the sample is real or ``generated" -- as in original GANs, we add the samples from GANs generator $G$ to the real $K$ classes data set, labeling them with a new ``generated" class $y=K+1$, so the classifier is expanded to the $K+1$ class. The loss function for training the classifier is defined as:
\begin{equation}\label{eg:GANsloss}
\begin{aligned}
L   = &-\E_{\mathbf{x},y \sim p_{d}(\mathbf{x},y)}[\log p_{m}(y|\mathbf{x})]  \\
 &-\E_{\mathbf{x} \sim G}[\log p_{m}(y=K+1|\mathbf{x})],
\end{aligned}
\end{equation}           
where we formulate two losses from the cross-entropy loss of the classifier as:

\vspace*{-5pt}
\begin{align}
 L_{sup} = &-\E_{\mathbf{x},y \sim p_{d}(\mathbf{x},y)}[\log p_{m}(y|\mathbf{x},{y<K+1})] \label{eq:GANsLsup},\\
 L_{unsup} = &-\{\E_{\mathbf{x}\sim p_{d}(\mathbf{x})}\log [1- p_{m}({y=K+1}|\mathbf{x}) \label{eq:GANsLunsup} \\
 &+ \E_{\mathbf{x} \sim G}\log [p_{m}(y=K+1|\mathbf{x})]\}.  \nonumber
\end{align}
 

\noindent Here, $\mathbf{x}$ stands for the given feature vector and $y$ is the corresponding label. The probability for the data and the model are defined as $p_{d}(.)$ and $p_{m}(.)$, respectively. In this way, we use all the labeled data to minimize the supervised loss, and we also employ unlabeled data in the unsupervised GANs training.

\section{Database Description}
\label{sec:database}
The dataset for the ADI task includes multi-dialectal speech from various programs recorded from Al-Jazeera TV channel. It includes audio files in MSA and four Arabic dialects: EGY, GLF, LAV, and NOR as well as their corresponding ASR transcripts and BNF i-Vector features. Table \ref{tab:datastat} presents some statistics about the training, development, and test datasets of MGB-3 ADI task \cite{Ali2017mgb3}.

\vspace{-6pt}
\begin{table}[h]\caption{The MGB-3 ADI databases: U stands for the number of utterances, D stands for the duration in hours, and W stands for the number of words in thousands.}\label{tab:datastat}
\vspace{5pt}
\centering
\scalebox{0.77}{
\begin{tabular}{|l||c c c|c c c|c c c|}
\hline
& \multicolumn{3}{c|}{\textbf{Training}} & \multicolumn{3}{c|}{\textbf{Development}} &  \multicolumn{3}{c|}{\textbf{Test}} \\
\textbf{Dialect} & \textbf{U} & \textbf{D} & \textbf{W} & \textbf{U} & \textbf{D} & \textbf{W} & \textbf{U} & \textbf{D} & \textbf{W} \\ \hline \hline
\textbf{EGY} & 3,093 & 12.4 & 76 & 298 & 2 & 11.0 & 302 & 2.0 & 11.6 \\ 
\textbf{GLF} & 2,744 & 10.0 & 56 & 264 & 2 & 11.9 & 250 & 2.1 & 12.3  \\
\textbf{LAV} & 2,851 & 10.3 & 53 & 330 & 2 & 10.3 & 334 & 2.0 & 10.9  \\
\textbf{MSA} & 2,183 & 10.4 & 69 & 281 & 2 & 13.4 & 262 & 1.9 & 13.0  \\
\textbf{NOR} & 2,954 & 10.5 & 38 & 351 & 2 & 9.9 & 344 & 2.1 & 10.3  \\ \hline \hline
\textbf{Total} & 13,825 & 53.6 & 292 & 1524 & 10 & 56.5 & 1492 & 10.1 & 58.1 \\ \hline 
\end{tabular}
}
\end{table}

\begin{figure}[htb]
\caption{2-dim LDA projection of the provided BNF i-Vector features for MGB-3 Corpus.}
\vspace{5pt}
\begin{minipage}[b]{.48\linewidth}
  \centering
  \centerline{\includegraphics[width=4.0cm]{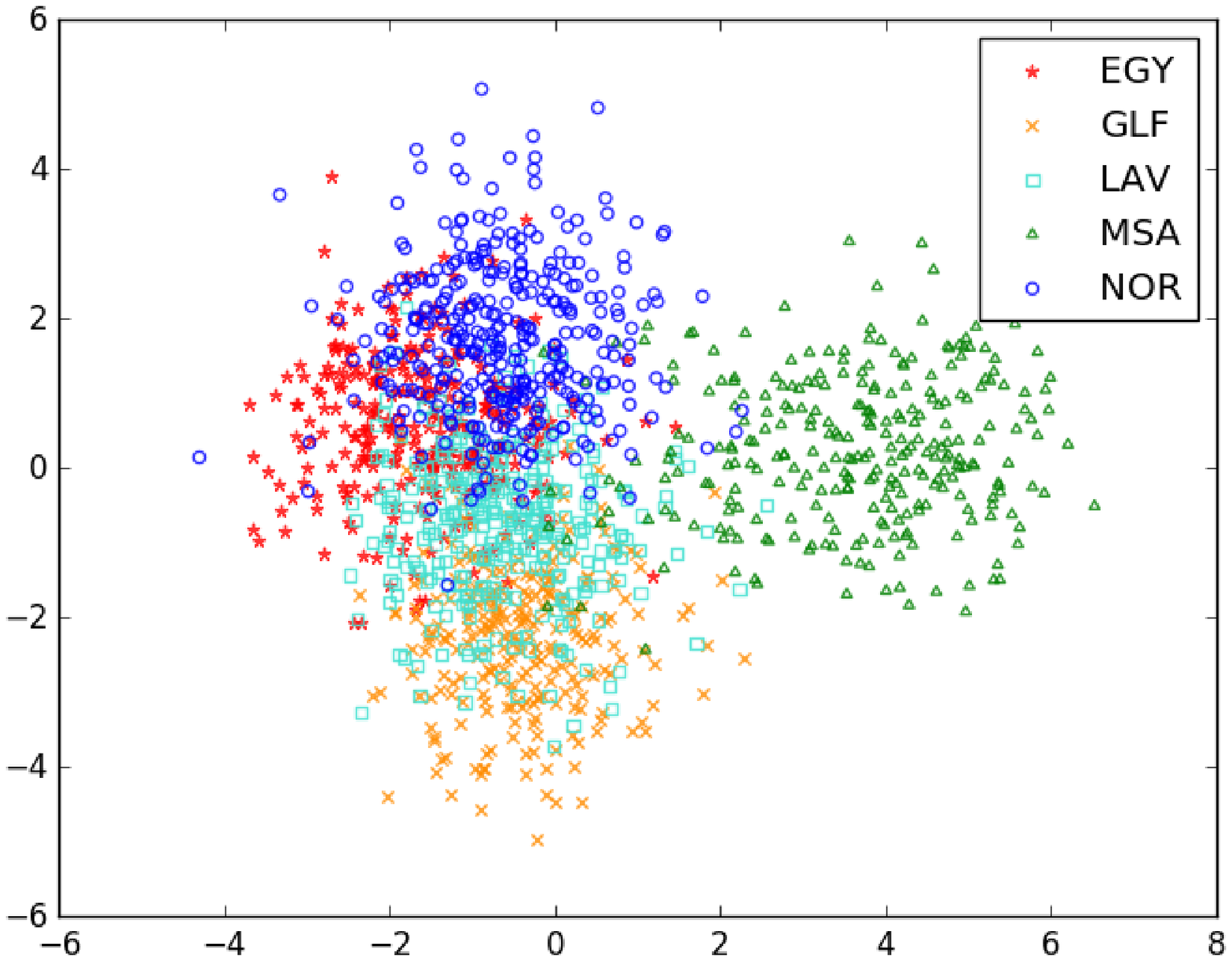}}
  \centerline{(b) Development Data}\medskip
\end{minipage}
\hfill
\begin{minipage}[b]{0.48\linewidth}
  \centering
  \centerline{\includegraphics[width=4.0cm]{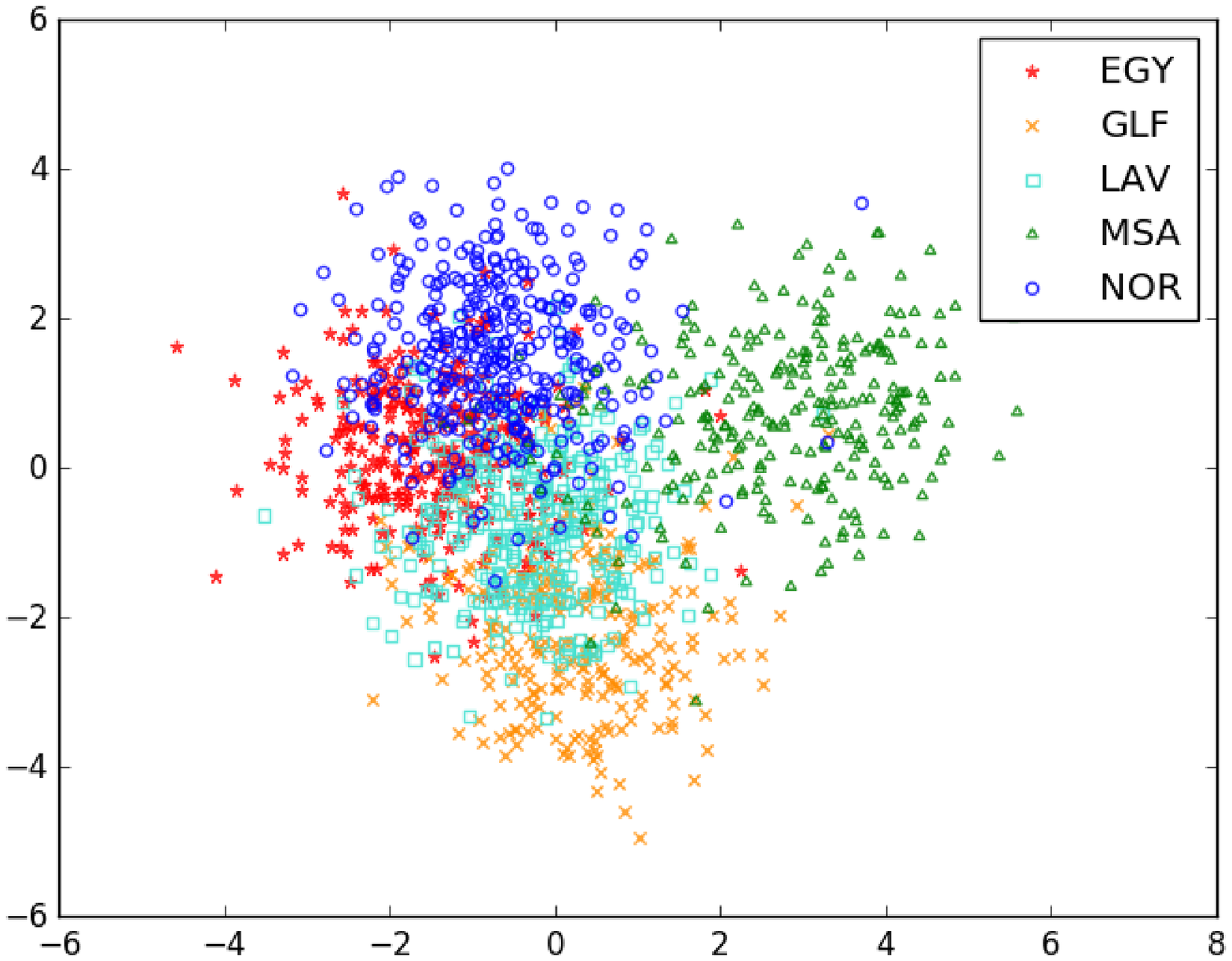}}
  \centerline{(c) Test Data}\medskip
\end{minipage}
\label{fig:bnfivecsdevtest}
\end{figure}

\noindent It is worth mentioning that the test and development datasets are collected from the same broadcast domain. Fig. \ref{fig:bnfivecsdevtest} depicts the 2-dim Linear Discriminant Analysis (LDA) projection of the provided BNF i-Vector features. It can be seen from the figure that the development and test data have similar class distributions. However, a different recording setup is built for the training dataset from a separate domain \cite{Ali2017mgb3}. 

Different from the previous ADI challenges, an additional 1,200 hours of training MGB-2 data is provided for this challenge. This data is also recorded from the various programs of the Al-Jazeera TV channel. Most data are estimated as MSA, and less than 30\% of the data is categorized as dialectal speech namely EGY, GLF, LAV, and NOR \cite{ali2016mgb}.

\section{Baseline Systems}
\label{sec:basesys}

The baseline system for the task uses lexical and BNF i-Vector features followed by a multi-class support vector machine (SVM) classifier \cite{AliDCKYG0R16}. We refer to those systems as \Romannum{1}\textsubscript{bs,t} and \Romannum{1}\textsubscript{bs,a}, respectively. In order to achieve a fair comparison with the baseline system, linear score fusion is employed for system combination \cite{brummer2007focal}, we refer to the fused system as \Romannum{1}\textsubscript{bs,a} + \Romannum{1}\textsubscript{bs,t}. Randomly selected one third of the development data is used for the fusion system training and 10-fold cross validation is applied. The overall performance of the baseline system is shown in Table \ref{tab:baseresults} using a 5-way classification setting.

\vspace{-8pt}
\begin{table}[h]\caption{The overall percentage accuracy, recall, and precision of the individual baseline text systems and fusion system on development set.}\label{tab:baseresults}
\vspace{5pt}
\centering
\begin{tabular}{|l||c c c|}
\hline
\textbf{System} & \textbf{ACC} & \textbf{RCL} & \textbf{PRC} \\ \hline \hline
\Romannum{1}\textsubscript{bs,t} & 48.26 & 50.33 & 49.13 \\ 
\Romannum{1}\textsubscript{bs,a} & 58.09 & 61.37 & 58.83  \\
\Romannum{1}\textsubscript{bs,a} + \Romannum{1}\textsubscript{bs,t} & \textbf{65.26} & \textbf{65.21} & \textbf{65.31}  \\ \hline
\end{tabular}
\end{table}

\section{Experiments and Results}
\label{sec:exps&results}

In this section, we describe experiments conducted and the results on both development and test datasets. Similar to baseline system described in the Sec. \ref{sec:basesys}, we randomly set aside one-third of the development data for the fusion training and 10-fold cross validation is applied for evaluation of development data performance. Once the best system configuration is determined on the development data it is fixed and applied to the test dataset. Throughout the experiments, the performance metric for ADI development and test datasets include Accuracy (ACC), Recall (RCL), and Precision (PRC).

\subsection{Text Based Systems}
\label{ssec:txsysperf}

As described in Sec. \ref{ssec:textsys}, 24 different text based systems are developed with all combinations of pre-processing methods and feature types. In Table \ref{tab:textsys}, the percentage accuracy of the best text based systems (the ones with over 50\% accuracy are reported) on the development dataset.

\begin{table}[h]\caption{Percentage accuracy of the best proposed text DID systems using the ADI MGB-3 development dataset.}\label{tab:textsys}
\vspace{5pt}
\centering
\begin{tabular}{|c|c|c|}
\hline
\textbf{Feature} & \textbf{Pre-Proc.} & \textbf{ACC (\%)} \\ \hline \hline
\multirow{3}{*}{BIN} & 4 & \textbf{51.11} \\ \cline{2-3}
& 1    & 50.72 \\ \cline{2-3}
& 3    & 50.65 \\ \hline
\multirow{2}{*}{TF} & 4 & \textbf{50.78} \\ \cline{2-3}
& 7    & 50.00 \\ \hline
\multirow{2}{*}{TF-IDF} & 1 & \textbf{50.52} \\ \cline{2-3}
& 3    & 50.00 \\ \hline
\end{tabular}
\end{table}


For each feature category, the best systems are chosen for the later back-end fusion of the systems. We experimentally realize that it is better to include best systems of each feature type rather than basically chose the best one overall. The systems for different feature types may have more complementary information even though they are not the best standalone system. It can be noted from Table \ref{tab:baseresults} that our reported proposed text based systems have superior performance versus the baseline text system (\Romannum{1}\textsubscript{bs,t}). We refer to the proposed text based systems as described in Table \ref{tab:textsysdesc}.

\vspace{-6pt}
\begin{table}[h]\caption{Description of the pre-processing methods and features for the proposed text based systems.}\label{tab:textsysdesc}
\vspace{5pt}
\centering
\begin{tabular}{|l||c c|}
\hline
\textbf{System} & \textbf{Pre-Proc.} & \textbf{Feature} \\ \hline \hline
\Romannum{1}\textsubscript{t} & 4 & BIN \\ 
\Romannum{2}\textsubscript{t} & 4 & TF  \\
\Romannum{3}\textsubscript{t} & 1 & TF-IDF \\ \hline
\end{tabular}
\end{table}

\subsection{Acoustic Based Systems}
\label{ssec:acosysperf}

As discussed in Sec. \ref{ssec:acusys}, we use three different types of features namely MFCC, BNF, and UBNF features for i-Vector extraction and two alternate back-ends namely GB and GANs for the classifier. The proposed acoustic based systems are summarized in Table \ref{tab:acousys}.

\vspace{-6pt}
\begin{table}[h]\caption{Description of the features and back-ends for the proposed acoustic based systems.}\label{tab:acousys}
\vspace{5pt}
\centering
\begin{tabular}{|l||c c|}
\hline
\textbf{System} & \textbf{Feature} & \textbf{Back-end} \\ \hline \hline
\Romannum{1}\textsubscript{a} & MFCC-ivecs & GB \\ 
\Romannum{2}\textsubscript{a} & MFCC-ivecs & GANs  \\
\Romannum{3}\textsubscript{a} & BNF-ivecs & GB \\ 
\Romannum{4}\textsubscript{a} & UBNF-ivecs & GANs  \\ \hline
\end{tabular}
\end{table}

In systems \Romannum{1}\textsubscript{a} and \Romannum{2}\textsubscript{a}, we use 20-dim MFCC vectors. The window length and shift size are 25-ms and 10-ms, respectively. Next, non-speech frames are discarded using energy-based voice activity detection. 1024-mixture full covariance UBM and total variability matrix have been trained using the whole MGB-2 and MGB-3 training dialectal data. Furthermore, 600-dim i-Vectors are extracted and classified by GB and GANs, respectively as described in Sec. \ref{sssec:classify}. It has to be noted that we apply LDA prior to GB but not to GANs.

For System \Romannum{3}\textsubscript{a}, 400-dim i-Vectors are used which are extracted using BNF features generated from an HMM-GMM baseline which is trained on 60 hours of manually transcribed Al-Jazeera MSA news recordings \cite{ali2016mgb}. In this system, we also use GB back-end classifier but no LDA projection is applied.

For System \Romannum{4}\textsubscript{a}, we use a very recent feature extraction scheme that has proven to be very beneficial for DID tasks \cite{zhang2017ubnf}. Firstly, a UBM model is trained with all MGB-3 training data based on MFCCs with Shifted Delta Cepstral (SDC) features. Specifically, the universal phonetic space is modeled with N Gaussian mixtures (i.e., N=2048 in our study). Subsequently, the frame level phonetic label is estimated according to posterior probability. There are only 4 hidden layers (1024-1024-40-2048) between the input and output layers because the size of DID training corpus in our study is around 50 hours. Furthermore, 600-dim i-Vectors are extracted using UBNF features and classified with GANs.






The details of the architecture of the GANs classifier used for Systems \Romannum{2}\textsubscript{a} and \Romannum{4}\textsubscript{a} are (100-500-500-601) for the generator and (601-1024-1024-1024-5) for the discriminator networks with 50\% random dropout for the hidden layers of the latter network.  We also append duration information to the i-Vector space, so that the input of the discriminator network and output of the generator network is 601 dimensions. For more details about how we employ GANs for semi-supervised training, and how duration information improves system performance, please see \cite{zhang2017letters}.


The results for the stand alone acoustic based systems are reported in Table \ref{tab:perfacusysall} for the development dataset with comparison to the baseline acoustic system.

\vspace{-6pt}
\begin{table}[h]\caption{The overall percentage accuracy, recall, and precision of the individual baseline and proposed acoustic systems on development set.}\label{tab:perfacusysall}
\vspace{5pt}
\centering
\begin{tabular}{|l||c c c|}
\hline
\textbf{System} & \textbf{ACC} & \textbf{RCL} & \textbf{PRC} \\ \hline \hline
\Romannum{1}\textsubscript{bs,a} & 58.09 & 61.37 & 58.83 \\ \hline
\Romannum{1}\textsubscript{a} & 63.37 & 63.82 & 64.29 \\
\Romannum{2}\textsubscript{a} & 60.78 & 61.92 & 61.46 \\
\Romannum{3}\textsubscript{a} & 58.82 & 62.98 & 59.24 \\
\Romannum{4}\textsubscript{a} & \textbf{69.42} & \textbf{68.97} & \textbf{69.83}  \\ \hline
\end{tabular}
\end{table}

\noindent The proposed acoustic based systems have clearly better performance than the baseline system. The first two proposed system performances show that classical MFCC i-Vectors have better performance for both back-end classifiers with sufficient amount of training data. The third system uses exactly the same features as the baseline system but has better performance with a GB classifier. As for the last proposed acoustic based system, we clearly see that it outperforms all other proposed and baseline systems. It is clear that UBNF based i-Vector features have a significantly better characterization of dialects for the ADI task.

In order to incorporate the proposed three text and four acoustic based systems, we use linear back-end score fusion \cite{brummer2007focal} as we employed for the baseline system. We carry out $2^7-1$ experiments on the development set in order to find the best fusion system combination among 4 acoustic and 3 text based systems. Table \ref{tab:perffusesyscombbest5} shows performances of the best five fusion combinations on the development set. Similar to the baseline system, the same randomly selected one-third of the development data is used for fusion system training and 10-fold cross validation is applied.

\vspace{-10pt}
\begin{table}[h]\caption{Overall percentage accuracy, recall, and precision of the best five fusion combinations over all systems on the MGB-3 development set.}\label{tab:perffusesyscombbest5}
\vspace{5pt}
\centering
\begin{tabular}{|l||c c c|}
\hline
\textbf{System Fusion Comb.} & \textbf{ACC} & \textbf{RCL} & \textbf{PRC} \\ \hline \hline
\Romannum{1}\textsubscript{a}+\Romannum{2}\textsubscript{a}+\Romannum{3}\textsubscript{a}+\Romannum{4}\textsubscript{a}+\Romannum{2}\textsubscript{t} & \textbf{75.97} & \textbf{75.82} & \textbf{76.26}  \\
\Romannum{1}\textsubscript{a}+\Romannum{3}\textsubscript{a}+\Romannum{4}\textsubscript{a}+\Romannum{2}\textsubscript{t} & 75.88 & 75.74 & 76.17 \\
\Romannum{1}\textsubscript{a}+\Romannum{2}\textsubscript{a}+\Romannum{3}\textsubscript{a}+\Romannum{4}\textsubscript{a}+\Romannum{1}\textsubscript{t}+\Romannum{2}\textsubscript{t} & 75.88 & 75.73 & 76.16  \\
\Romannum{1}\textsubscript{a}+\Romannum{2}\textsubscript{a}+\Romannum{3}\textsubscript{a}+\Romannum{4}\textsubscript{a}+\Romannum{1}\textsubscript{t}+\Romannum{2}\textsubscript{t}+\Romannum{3}\textsubscript{t} & 75.69 & 75.54 & 76.00 \\ 
\Romannum{1}\textsubscript{a}+\Romannum{2}\textsubscript{a}+\Romannum{3}\textsubscript{a}+\Romannum{4}\textsubscript{a}+\Romannum{1}\textsubscript{t} & 75.23 & 75.06 & 75.47 \\ \hline
\end{tabular}
\end{table}

\begin{figure}[htb]
\caption{The overall percentage accuracy, recall, and precision of step-by-step best fusion combination on development set with comparison to the fused baseline system.}
\label{fig:perffusesysvsbase}
\includegraphics[width=0.96\linewidth]{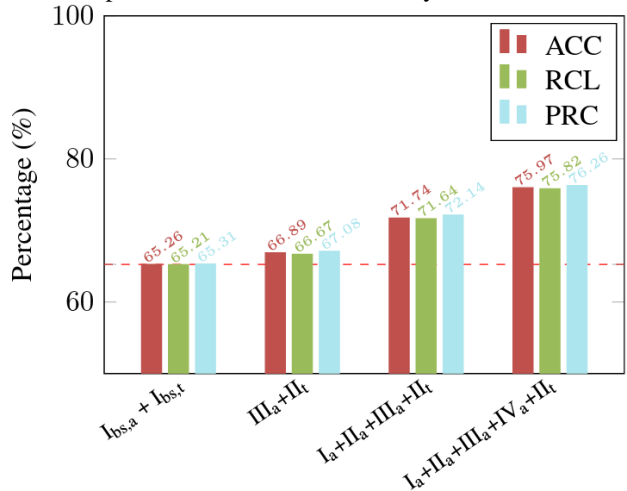}
\end{figure}

    \noindent Here, we point out that the second best-proposed text system makes the leading contribution in the fused system. Also, it is not always the case that combining the best systems at hand, achieves the best performance overall. All proposed acoustic based systems clearly contribute to improved fusion system with different amounts of gain. In Fig. \ref{fig:perffusesysvsbase}, the contributions of the each system can be seen with comparison to the fused baseline system.


\vspace{-12pt}
\vspace{-8pt}

\noindent The baseline fusion system is shown in the very left of the Fig. \ref{fig:perffusesysvsbase}. As a first proposed system, we develop one acoustic and one text based fusion system. As described in the Table \ref{tab:acousys}, our first fused system uses acoustic and text features and has minor accuracy improvement over baseline system. Next, the systems using MFCC i-Vectors are incorporated into the fused systems and a further 5\% absolute accuracy improvement is achieved. Lastly, the system that uses the UBNF i-Vectors is added to the combination of the four systems and the best accuracy is observed with a 4.2\% absolute accuracy increase on the development set. It can be concluded that each feature type has a great deal of complementary information for the ADI task.

\section{Submissions}
\label{sec:submissions}

After release of the test data, in accordance with the analysis on development dataset, we made the ``\texttt{primary}" submission with the best fused system, \Romannum{1}\textsubscript{a}+\Romannum{2}\textsubscript{a}+\Romannum{3}\textsubscript{a}+\Romannum{4}\textsubscript{a}+\Romannum{2}\textsubscript{t} as shown in Fig. \ref{fig:perffusesysvsbase}. In order to train back-end classifiers, we use the MGB-3 training set, and for training the fusion model of the system scores we use the entire development dataset. With the reported ``\texttt{primary}" performance results as shown in Table \ref{tab:subperf}, we were ranked as the second best system for the MGB-3 ADI task. 

As pointed out in Sec. \ref{sec:database}, the development and test datasets are collected from the same stream with the same setup. Class distributions of the development and test datasets can be seen in Fig. \ref{fig:bnfivecsdevtest}. In order to leverage the similarity of the development data with test data, we augment randomly selected two thirds of the development data to the training data for training the back-end classifiers. The second system is submitted as ``\texttt{contrastive2}" to the challenge. Performance comparison can be seen in Table \ref{tab:subperf}.

The importance of the inclusion of development set in back-end training is significant as shown in Table \ref{tab:subperf}. The relative performance increase is around 9\% with just this data augmentation.

After the test key was released, we realize that we had problems with the MFCC i-Vector based systems on test set namely \Romannum{1}\textsubscript{a} and \Romannum{2}\textsubscript{a}. For the systems, there was a considerable performance mismatch as opposed to the development set. After the feature extraction problem of the systems is fixed, we re-ran the best-fused system on the test set and reported the results as post-evaluation analysis in Table \ref{tab:subperf}.

\vspace{-7pt}
\begin{table}[h]\caption{The performance of the system submissions on MGB-3 test set with different set combinations for back-end and system fusion trainings.}\label{tab:subperf}
\vspace{5pt}
\centering
\scalebox{0.82}{
\renewcommand{\arraystretch}{1.2}
\begin{tabular}{|c||c c||c c c|}
\hline
\textbf{Submission} & \textbf{Back-end} & \textbf{Fusion} &\textbf{ACC} & \textbf{RCL} & \textbf{PRC} \\ \hline \hline
\texttt{primary} & train & dev & 70.38 & 70.78 & 71.69  \\ 
\texttt{contrastive2} & train + $\frac{2}{3}$ dev & $\frac{1}{3}$ dev & 76.94 & 76.95 & 77.60 \\
\texttt{post\_eval} & train + $\frac{2}{3}$ dev & $\frac{1}{3}$ dev & \textbf{79.76} & \textbf{79.87} & \textbf{80.27} \\ \hline
\end{tabular}
}
\end{table}

\noindent To the extent of our knowledge, systems \texttt{contrastive2} and \texttt{post\_eval} have the best performance on the MGB-3 ADI test data as compared to the results in \cite{zampieri2017vardial17, Ali2017mgb3}.

\section{Conclusion}
\label{sec:conclude}

In this study, we successfully combine four acoustic based systems with a text based system for the MGB-3 ADI challenge. MFCC, BNF, and UBNF based i-Vector features are used for the acoustic based systems with GB and GANs back-end classifiers. For the text based system, TF based unigram features are employed with an SVM classifier. It was shown that the combination of those five systems achieves very successful results on the test set with back-end score fusion.

\bibliographystyle{IEEEbib}
\bibliography{strings,refs}

\end{document}